\documentclass[onecolumn,amsmath,showkeys,amssymb]{revtex4}

\usepackage{graphicx}

\usepackage{graphicx}
\usepackage{color}

\usepackage{amsbsy}
\usepackage{amsthm,mathrsfs,amsopn}

\begin{document}

\title{Pattern formation for reactive species undergoing anisotropic diffusion}
\author{Daniel M. Busiello$^{1}$, Gwendoline Planchon$^{2}$, Malbor Asllani$^2$, Timoteo Carletti$^2$ and Duccio Fanelli$^3$}

\affiliation{1. Dipartimento di Fisica e Astronomia "G. Galilei" Via Marzolo, 8, I-35131, Padova, Italy\\
2. Department of mathematics and Namur Center for Complex Systems - naXys, University of Namur, rempart de la Vierge 8, B 5000 Namur, Belgium\\
3. Dipartimento di Fisica e Astronomia, University of Florence,INFN and CSDC, Via Sansone 1, 50019 Sesto Fiorentino, Florence, Italy\\
}

\begin{abstract} 
Turing instabilities for a two species reaction-diffusion systems is studied under anisotropic diffusion. More specifically, the diffusion constants which characterize the ability of the species to relocate in space are direction sensitive. Under this working hypothesis, the conditions for the onset of the  instability are mathematically derived and numerically validated. Patterns which closely resemble those obtained in the classical context of isotropic diffusion, develop when the usual Turing condition is violated, along one of the two accessible directions of migration. Remarkably, the instability can also set in when the activator diffuses faster than the inhibitor, along the direction for which the usual Turing conditions are not matched.   
\end{abstract}

\maketitle

\vspace{0.8cm}

\section{Introduction}
\label{sec:intro}

Spatio-temporal patterns are widespread in nature: beautiful spots and stripes appear on the coat of animals~\cite{Murray2}, patterns of cracking emerge on the fracture surface of materials~\cite{Ord2010}, reacting chemicals give rise to complex and dynamical structures as in the celebrated Belousov-Zhabotinsky reaction~\cite{Belousov, Strogatz,Yang}, spatial games in social sciences yield self-organized regular motifs~\cite{Novak1992,Novak1993,Wakano2011,Deforest2013}. A common feature which is shared by the above mentioned applications is the spontaneous formation of complex structures, which result from the non trivial interplay between noise and deterministic dynamics. Elucidating the key mechanisms that seed the process of pattern formation is therefore an important topic of investigation of cross disciplinary impact.

One of such mechanisms was identified and thoroughly discussed in a pioneering work of A. Turing~\cite{turing1952}: homogeneous equilibrium solutions of a multi-species reaction-diffusion system can be destabilized upon injection of a small inhomogeneous perturbation. This latter undergoes an exponential amplification, in the linear regime of the evolution. Then, non linearities come into play and the system eventually reaches a patchy, spatially inhomogeneous, equilibrium. Traveling waves and spiraling patterns can be also generated following a Turing-like, symmetry breaking instability. 

In the classical setting, two mutually interacting species are considered: these are the so called activator and  inhibitor.  
If the diffusion is isotropic, or in other words not affected by the specific direction of {displacement}, the inhibitor species should diffuse faster than the activator, for Turing patterns to develop. Systems of three simultaneously diffusing species ~\cite{satnoianu} have also been considered in the literature and shown to display a richer zoology of possible instabilities and pattern. In this generalized context, self-organized motifs can also develop, if one species is solely allowed to diffuse in the embedding medium~\cite{ErmentroutLewis1997}. Beyond the deterministic scenario, stochastic Turing patterns have been also reported for reaction diffusion systems defined on a regular lattice or complex networks \cite{Biancalani2010,Cantini2013,Malbor2012,Malbor2013}.  

Starting from these premises, and with reference to the paradigmatic scenario where just two species are made to interact, we shall here revisit the conditions that yield the Turing instability, under the assumption of anisotropic diffusion. More concretely, we shall derive sufficient conditions for the emergence of Turing like patterns in a rectangular, continuous, domain subject to periodic boundary conditions, assuming generic non linear reaction terms and imposing anisotropic, i.e. direction sensitive, diffusion coefficients. As we will demonstrate in the following, patterns do exist also if the condition for the onset of the Turing instability is uniquely satisfied along one direction. These latter patterns resemble quite closely those that are found under the standard assumption of isotropic diffusion,  {the non linearity being responsible for the mixing of cross modes}. In addition, patterns can also flourish when the activator diffuses faster than the inhibitor, along one specific direction. In this case, the system organizes along the direction orthogonal to the latter, hence displaying regular, just locally distorted, stripes. 

The paper is organized as follows. In Section~\ref{sec:contdom} we will present the reference framework and then, in Section \ref{ssec:hsign} derive the mathematical conditions for the generalized anisotropic instability. Section~\ref{sec:applications} is devoted to reporting some numerical tests to validate the theoretical analysis. Finally, we shall sum up and conclude.

\section{Anisotropic diffusion of reactive species on continuum domains}
\label{sec:contdom}

Let us consider two interacting species and denote by $u$ and $v$ their respective concentrations. The species can freely diffuse inside a rectangular
domain, $R = [0,L_x]\times [0,L_y] \subset \mathbb{R}_+\times \mathbb{R}_+$, as specified by their respective diffusion coefficients. We shall in particular assume
that the diffusion coefficients are anisotropic, meaning that they depend on the specific direction of migration. More precisely, $D_u^{(x)}\geq 0$ denotes the diffusion coefficient for species $u$ along direction $x$, while $D_u^{(y)}\geq 0$ refers to the orthogonal direction $y$. Similar considerations respectively apply to $D_v^{(x)}\geq 0$ and $D_v^{(y)}\geq 0$. The mutual evolution of species $u$ and $v$ is thus governed by the reaction diffusion equations:  
\begin{equation}
 \label{eq:contmodPDE}
 \begin{cases}
 \dot{u}&=f(u,v)+D_u^{(x)} \partial^2_x u+D_u^{(y)} \partial^2_y u\\
 \dot{v}&=g(u,v)+D_v^{(x)} \partial^2_x v+D_v^{(y)} \partial^2_y v
 \end{cases}\quad \forall (x,y)\in R \quad \text{and}\quad \forall t>0 \, .
\end{equation}
where $f(\cdot,\cdot)$ and $g(\cdot,\cdot)$ are non linear functions of the concentration amounts.
The above equations should be complemented by the initial conditions:
\begin{equation}
 \label{eq:contmodinicond}
u(x,y,0)=u_0(x,y) \quad \text{and} \quad v(x,y,0)=v_0(x,y) \quad \forall (x,y)\in R\, ,
\end{equation}
for some regular functions $u_0$ and $v_0$, and suitable boundary conditions. In the following we shall adopt the Dirichlet periodic boundary conditions, namely
\begin{equation}
 \label{eq:contmodBC}
  \begin{cases}
u(x,0,t)=u(x,L_y,t) \quad \forall x\in [0,L_x] \quad\text{and} \quad\forall t>0\\
u(0,y,t)=u(L_x,y,t) \quad \forall y\in [0,L_y]\quad \text{and} \quad\forall t>0
  \end{cases}\, ,
\end{equation}
and similarly for $v$.

Let us assume the system~\eqref{eq:contmodPDE} admits a stable, spatially homogeneous, solution $u=\hat{u}$ and $v=\hat{v}$. This request translates in:
\begin{equation}
 \label{eq:contmodEquil}
  \begin{cases}
f(\hat{u},\hat{v})=0\\
g(\hat{u},\hat{v})=0
  \end{cases}\, \text{such that:} \quad\mathrm{tr}(J)=f_u+g_v<0\quad \text{and}\quad  \det(J)=f_u g_v-f_v g_u>0\, ,
\end{equation}
where $J$ stands for the Jacobian matrix of system (\ref{eq:contmodPDE}):
\begin{equation}
J=\left( \begin{array}{cc}
f_u & f_v \\
g_u & g_v \\
\end{array} \right)\
\end{equation}
where $f_u$ denotes the derivative of $f(u,v)$ with respect to $u$, and similarly for $f_v,g_u,g_v$. Here, and throughout the remaining part of the paper, we evaluate the partial derivatives at the equilibrium point $(\hat{u},\hat{v})$. Without losing generality, we will also assume $f_u>0$ and $g_v<0$: $u$ is thus the activator species, while $v$ refers to the population of inhibitors.

The celebrated Turing patterns originate from a symmetry breaking instability of the homogeneous equilibrium solution. The introduction of an inhomogeneous perturbation around $(\hat{u},\hat{v})$ activates the diffusion terms and, under specific conditions, makes the system to drift away from the deputed homogeneous equilibrium, towards a patchy, non homogeneous, asymptotically stable, solution. Mathematical conditions for the Turing instability to set in can be readily derived by first linearizing equations~\eqref{eq:contmodPDE} and then Fourier transforming, both in time and space, the obtained linear system. This yields the so called dispersion relation, an equation for the growth rate $\lambda_k$ associated to Fourier mode $k=(k_x,k_y)$. By carrying out this straightforward calculation, which is for instance detailed in \cite{Murray2}, it can be eventually proven that $\lambda_k$ satisfies the following quadratic equations:
\begin{equation}
\label{eq:reldisp}
\lambda_k^2+d(k_x,k_y)\lambda_k+h(k_x,k_y)=0\, , 
\end{equation}
where:
\begin{eqnarray}
\label{eq:kxhx}
d(k_x,k_y)&=&-\mathrm{tr}(J)+k_x^2(D_u^{(x)}+D_v^{(x)})+k_y^2(D_u^{(y)}+D_v^{(y)})\\
h(k_x,k_y)&=&\det(J)-k_x^2(f_u D_v^{(x)}+g_vD_u^{(x)})-k_y^2(f_uD_v^{(y)}+g_vD_u^{(y)})\label{eq:hkxhx}\\&+&k_x^2k_y^2(D_u^{(x)}D_v^{(y)}+D_u^{(y)}D_v^{(x)})+k_x^4 D_u^{(x)}D_v^{(x)}+k_y^4D_u^{(y)}D_v^{(y)}\notag\, .
\end{eqnarray}

Turing patterns materialize if the real part of $\lambda_k$ takes positive values over finite window in $k$, which in turn amounts to require the presence of unstable non zero Fourier modes. We remark however that $d(k_x,k_y)$ in Eq.~\eqref{eq:reldisp} is always positive, since, by assumption, $\mathrm{tr}(J)<0$ and, in addition, $D_{(u,v)}^{(x),(y)}>0$. Then, as a natural consequence, the Turing symmetry breaking instability can take place only if a compact domain exists in $(k_x,k_y)$ such that $h(k_x,k_y)<0$. As already mentioned, in the classical limit of isotropic diffusion,  $D_u \equiv D_u^{(x)}=D_u^{(y)}$ and $D_v \equiv D_v^{(x)}D_v^{(x)}$, the Turing instability can take place only if the inhibitors diffuse faster than the activators, i.e. {$D_v > r_cD_u$ where $r_c$, the critical ratio of diffusivities, is a positive coefficient larger than $1$}. In the following we will show that this stringent assumption can be partially relaxed in the generalized setting where the diffusion constants are made to depend on the direction of propagation.

\section{Turing instability in presence of anisotropic diffusion}
\label{ssec:hsign}

The function $h(k_x,k_y)$ is a multivariate polynomial of the variables $k_x^2$ and $k_y^2$. It is straightforward to check that it is positive at the origin and for large $k_x^2$ and $k_y^2$. We are here interested in determining when $h(k_x,k_y)$ can change sign as function of $k_x^2$ and $k_y^2$, so signaling the onset of the instability. To this end, we first consider restrictions of $h(k_x,k_y)$ on $k_x=0$, and then on $k_y=0$.

Focusing on the restriction of $h$ on the $k_y$ axis, i.e. namely setting $k_x=0$, amounts to consider the particular case where species $u$ and $v$ are solely allowed to diffuse along the vertical direction. One can therefore equivalently set $D_u^{(x)}=D_v^{(x)}=0$ in Eq~\eqref{eq:hkxhx} and thus get:
\begin{equation}
\label{eq:kx0}
h(k_x,k_y)=\det(J)-k_y^2(f_u D_v^{(y)}+g_vD_u^{(y)})+k_y^4D_u^{(y)}D_v^{(y)}\, ,
\end{equation}
By solving equation (\ref{eq:kx0}) for $k_y^2$, one obtains two positive solutions, $0<k_-<k_+$, if and only if the following conditions are met:
\begin{equation}
\label{eq:case1}
\begin{cases}
f_u D_v^{(y)}+g_v D_u^{(y)}>0\\
(f_u D_v^{(y)}+g_vD_u^{(y)})^2-4D_u^{(y)}D_v^{(y)} \det(J) >0\, .
\end{cases}
\end{equation}

Let us observe that from the first relation of Eq.~\eqref{eq:case1} and the condition $\mathrm{tr}(J)<0$ implies $D_v^{(y)}>D_u^{(y)}$: for the instability to set in and the patterns to develop, the inhibitor should diffuse faster than the activator in the $y$ direction.

The symmetric limiting case is recovered when species $u$ and $v$ are allowed to diffuse only along the horizontal direction, which in turn amounts to restrict $h$ to the $k_x$ axis. The analysis can be hence handled by setting $D_u^{(y)}=D_v^{(y)}=0$ in Eq~\eqref{eq:hkxhx} and proceeding in analogy with above. One can straightforwardly obtain the following necessary and sufficient conditions for the existence of Turing patterns:
\begin{equation}
\label{eq:case2}
\begin{cases}
f_u D_v^{(x)}+g_vD_u^{(x)}>0\\
(f_u D_v^{(x)}+g_v D_u^{(x)})^2-4D_u^{(x)}D_v^{(x)} \det(J) >0\, .
\end{cases}
\end{equation}
Once again, from the first relation of Eq.~\eqref{eq:case2} and the condition $\mathrm{tr}(J)<0$, one can immediately conclude that patterns are possible only if 
$D_v^{(x)}>D_u^{(x)}$, namely if the inhibitor diffuses faster than the activator along the $x$ direction.

These conclusions are clearly not surprising, as they constitute an obvious adaptation of the standard Turing framework to the present context, in the trivial limit where one of the diffusion direction is alternatively silenced. Starting from this observation, it is however interesting to speculate on the possibility of turning unstable complex mixed modes ($k_x,k_y$), via a symmetry breaking process of the Turing type, when the simplified pathways to pattern formation explored above are instead precluded. 

To this end, we go back to function $h(k_x,k_y)$ and study its sign when moving on ($k_x,k_y$), along specific directions. More concretely, we set  $k_x=\gamma k_y$, and vary the free parameter $\gamma$ to span the reference plane. Turing patterns can then develop only if $h(\gamma k_y,k_y)<0$, where:
\begin{eqnarray}
\label{eq:kxkyApp}
h(\gamma k_y,k_y)&=&\det(J)-k_y^2[\gamma^2(f_u D_v^{(x)}+g_vD_u^{(x)})+(f_uD_v^{(y)}+g_vD_u^{(y)})]\notag\\&+&k_y^4[\gamma^2(D_u^{(x)}D_v^{(y)}+D_u^{(y)}D_v^{(x)})+\gamma^4 D_u^{(x)}D_v^{(x)}+D_u^{(y)}D_v^{(y)}]\notag\\
&=:&B_1 k_y^4-B_2k_y^2+B_3 \, ,
\end{eqnarray}
and the last expression defines the coefficients $B_1$, $B_2$ and $B_3$. It can be readily realized that $B_1$ and $B_3$ are positively definite, while $B_2$ can assume both positive and negative values. In the following, we shall impose the simultaneous violation of conditions~\eqref{eq:case1} and~\eqref{eq:case2}, via crossed negation of the corresponding inequalities, and look for possible values of the control parameter $\gamma$ that make the system unstable.

\subsection{Conditions~\eqref{eq:case1}i and \eqref{eq:case2}i are not satisfied}
Let us thus assume
\begin{equation*}
\begin{cases}
f_u D_v^{(y)} + g_v D_u^{(y)} <0\\
f_u D_v^{(x)} + g_v D_u^{(x)} <0\, ,
\end{cases}
\end{equation*}
while the remaining two conditions\eqref{eq:case1}ii and~\eqref{eq:case2}ii do hold.

One can trivially realize that in this case $B_2$ is negative, hence $h(\gamma k_y,k_y)=B_1 k_y^4+\lvert B_2\rvert k_y^2+B_3>0$ for all $k_x=\gamma k_y$ and $k_y$. No instability can thus develop which seeds the emergence of self-organized Turing patterns.  

\subsection{Conditions~\eqref{eq:case1}i and \eqref{eq:case2}ii are not satisfied}

We now assume
\begin{equation*}
\begin{cases}
f_u D_v^{(y)} + g_v D_u^{(y)} <0\\
(f_u D_v^{(x)} + g_v D_u^{(x)})^2-4D_u^{(x)}D_v^{(x)} \det(J)<0\, ,
\end{cases}
\end{equation*}
while the remaining two relations \eqref{eq:case1}ii and~\eqref{eq:case2}i are verified.

Solving for the limiting condition $h(\gamma k_y,k_y)=0$ one gets  a closed expression for $k_y^2$. By imposing $k_y^2$ to be positive yields 
$B_2>0$ and $B_2^2-4B_1B_3>0$.

A straightforward computation gives:
\begin{equation*}
B_2>0\quad\text{if $\gamma^2>q_1$,}
\end{equation*}
where 
\begin{equation*}
q_1=-\frac{f_u  D_v^{(y)} + g_v D_u^{(y)}}{f_u D_v^{(x)} + g_v D_u^{(x)}}>0\, ,
\end{equation*}
where use has been made of Eq.~\eqref{eq:case2}i.

A somehow lengthy computation allows us to write:
\begin{equation}
\label{eq:gamma1}
B_2^2-4B_1B_3=A_1  \gamma^4  + A_2 \gamma^2 +A_3 \, ,
\end{equation}
where:
\begin{eqnarray}
\label{eq:Ai}
A_1 &=& \Gamma_1 - 4\det(J)D_u^{(x)} D_v^{(x)}\\
A_2 &=& 2(f_u D_v^{(x)} + g_v D_u^{(x)}) (f_u D_v^{(y)} + g_v D_u^{(y)}) - 4\det(J)(D_u^{(y)}D_v^{(x)} + D_u^{(x)}D_v^{(y)})\\
A_3 &=& \Gamma_2 - 4\det(J)D_u^{(y)} D_v^{(y)}
\end{eqnarray}
and
\begin{equation}
\label{eq:gammai}
\Gamma_1 = (f_u D_v^{(x)} + g_v D_u^{(x)})^2\quad \text{and}\quad  \Gamma_2 = (f_u D_v^{(y)} + g_v D_u^{(y)})^2\, .
\end{equation}

Under the above assumptions $\Gamma_1 < 4\det(J)D_u^{(x)} D_v^{(x)}$, which implies $A_1<0$. Similarly, as $\Gamma_2 > 4\det(J)D_u^{(y)} D_v^{(y)}$,  $A_3>0$. On the other hand, $A_2<0$, this latter quantity resulting from the sum of two negative terms. Hence, $B_2^2-4B_1B_3>0$ if $0<\gamma^2<q_2$, where $q_2=\frac{A_2+\sqrt{A_2^2-4A_1A_3}}{-2A_1}>0$.

We can easily show that $q_1>q_2$, which in turn implies that $B_2$ and $B_2^2-4B_1B_3$ cannot be at the same time positive, as it should happen for the instability to develop. We can hence conclude that Turing patterns cannot develop in this case either.

\subsection{Conditions~\eqref{eq:case1}ii and \eqref{eq:case2}i are not satisfied}

Let us thus assume
\begin{equation*}
\begin{cases}
(f_u D_v^{(y)} + g_v D_u^{(y)})^2-4D_u^{(y)}D_v^{(y)} \det(J)<0\\
f_u D_v^{(x)} + g_v D_u^{(x)} <0\, ,
\end{cases}
\end{equation*}
while the remaining two condition\eqref{eq:case1}i and~\eqref{eq:case2}ii are verified.

Once again requiring $h(\gamma k_y,k_y)<0$, necessarily imply
$B_2>0$ and $B_2^2-4B_1B_3>0$. The former condition is satisfied whenever:
\begin{equation*}
\gamma^2 \in (0,q_1)\, ,
\end{equation*}
for $q_1=-(f_u D_v^{(y)}+g_v D_u^{(y)})/(f_u D_v^{(x)}+ g_v D_u^{(x)})>0$. The latter condition $B_2^2-4B_1B_3>0$ yields
\begin{eqnarray*}
&\gamma^4& (\Gamma_1-4\det(J)D_u^{(x)}D_v^{(x)})+\gamma^2[2(f_u D_v^{(x)}+g_v D_u^{(x)})(f_u D_v^{(y)}+g_v D_u^{(y)})-4\det(J)(D_u^{(y)}D_v^{(x)}+D_u^{(x)}D_v^{(x)})\\&+&\Gamma_2-4\det(J)D_u^{(y)}D_v^{(y)}:=\gamma^4 A_1+\gamma^2A_2+A_3>0\, .
\end{eqnarray*}

Here, $A_1>0$ while $A_2<0$ and $A_3<0$. Hence, the previous inequality is satisfied for any $\gamma^2>q_2$ for $q_2=(-A_2+\sqrt{A_2^2-4A_1A_3})/(2A_1)>0$. However, one can prove that $q_1<q_2$, which implies that $B_2$ and $B_2^2-4B_1B_3$ cannot be simultaneously positive. The conclusion is therefore that  $h(\gamma k_y,k_y)>0$, and Turing patterns cannot take place.

\subsection{Conditions~\eqref{eq:case1}ii and \eqref{eq:case2}ii are not satisfied}

Let us thus assume
\begin{equation*}
\begin{cases}
(f_u D_v^{(y)} + g_v D_u^{(y)})^2-4D_u^{(y)}D_v^{(y)} \det(J)<0\\
(f_u D_v^{(x)} + g_v D_u^{(x)})^2-4D_u^{(x)}D_v^{(x)} \det(J)<0\, ,
\end{cases}
\end{equation*}
while the remaining two assumptions~\eqref{eq:case1}i and~\eqref{eq:case2}i do hold.

Under the present working hypothesis, the coefficient $B_1$, $B_2$ and $B_3$ are positive. Thus $h(\gamma k_y , k_y)$ can take negative values, if and only if $B_2^2-4B_1 B_3 >0$. As previously remarked, we can rewrite
\begin{equation*}
B_2^2-4B_1B_3=A_1  \gamma^4  + A_2 \gamma^2 +A_3 \, ,
\end{equation*}
where $A_i$ for $i=1,2,3$ are defined as in~\eqref{eq:Ai}. One can show that $A_1$ and $A_3$ are negative while $A_2$ can take both signs. To satisfy the requirement $B_2^2-4B_1 B_3 >0$ the conditions $A_2>0$ and $A_2^2-4A_1A_3>0$ should be simultaneously met.

Let us rewrite $A_2$ as follows
\begin{equation*}
A_2 = 2\sqrt{\Gamma_1} \sqrt{\Gamma_2} - 4\det(J)\left(D_u^{(y)} D_v^{(x)}+D_u^{(x)}D_v^{(y)}\right)\, ,
\end{equation*}
where $\Gamma_i$ have been defined in Eq.~\eqref{eq:gammai}. Straightforward manipulations allow us to write:
\begin{eqnarray*}
A_2 &=& 2\sqrt{\Gamma_1} \sqrt{\Gamma_2} - 4\det(J)\left(D_u^{(y)} D_v^{(y)} \frac{D_v^{(x)}}{D_v^{(y)}}+D_u^{(x)} D_v^{(x)}\frac{D_v^{(y)}}{D_v^{(x)}}\right)\\&<& 2\sqrt{\Gamma_1} \sqrt{\Gamma_2} - \left(\Gamma_2\frac{D_v^{(x)}}{D_v^{(y)}}+\Gamma_1\frac{D_v^{(y)}}{D_v^{(x)}}\right)
=-\left(\sqrt{\frac{D_v^{(y)}}{D_v^{(x)}}\Gamma_1}-\sqrt{\frac{D_v^{(x)}}{D_v^{(y)}}\Gamma_2}\right)^2<0\, .
\end{eqnarray*}
Since $A_2$ is bound to be negative, the condition for Turing instability $h(\gamma k_y , k_y)<0$ cannot be satisfied.

\begin{figure}[htbp!]
\centering
\includegraphics[width=18cm]{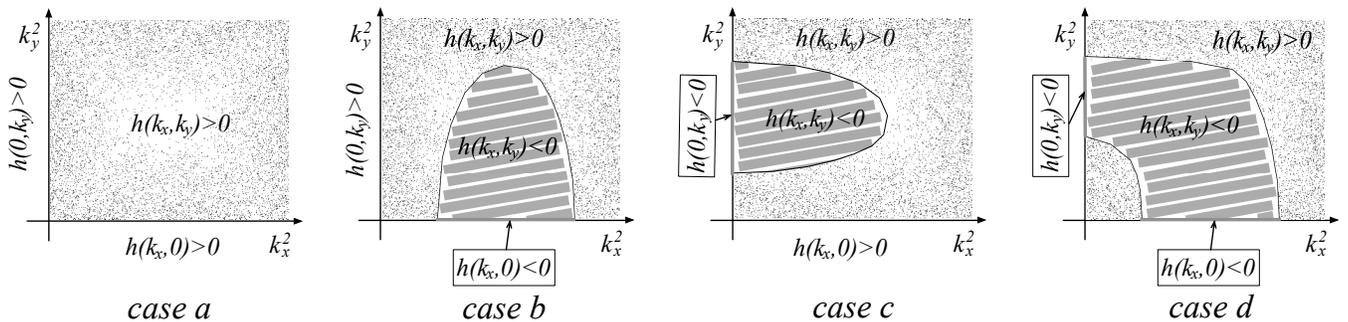}
\caption{Possible types of instabilities. Case a: $h(k_x,k_y)>0$ for all $k_x^2\geq 0$ and $k_y^2\geq 0$. The system cannot turn unstable. Case b: $h$ restricted to the $k_x$ axis takes negative values: a bounded contiguous domain in $k_x^2> 0$ and $k_y^2> 0$ exist, for which $h(k_x,k_y)>0$. Case c: $h$ restricted to the $k_y$ axis takes negative values. Again a portion of the reference plan, adjacent to the domain of instability in $k_x=0$, can be found where $h(k_x,k_y)>0$: Case d: the system is unstable along both $k_x=0$ and $k_y=0$ directions. The instability also interests non trivial modes with both $k_x \ne 0$ and $k_y \ne = 0$.
}
\label{fig:hsign}
\end{figure}

Summing up we have demonstrated that patterns can eventually develop only if the system can undergo a symmetry breaking instability of the Turing type, 
in its restricted configuration where the diffusion is solely allowed along one spatial direction, either $x$ or $y$.  The result is summarized in Figure~\ref{fig:hsign}, where different types of instabilities are schematically depicted. 

Interestingly, the instability can set in also if the inhibitor diffuses slower that the activator along one selected direction, provided the opposite holds for the transport along the orthogonal direction. In this respect, accounting for anisotropic diffusion enables one to partially relax the stringent conditions that underly the formation of the Turing motifs. In the next section, we will built on this observation and provide a numerical demonstration of the investigated phenomenon.

\section{Numerical analysis}
\label{sec:applications}

\begin{figure}[htbp!]
\begin{tabular}{cc}
\includegraphics[width=6cm]{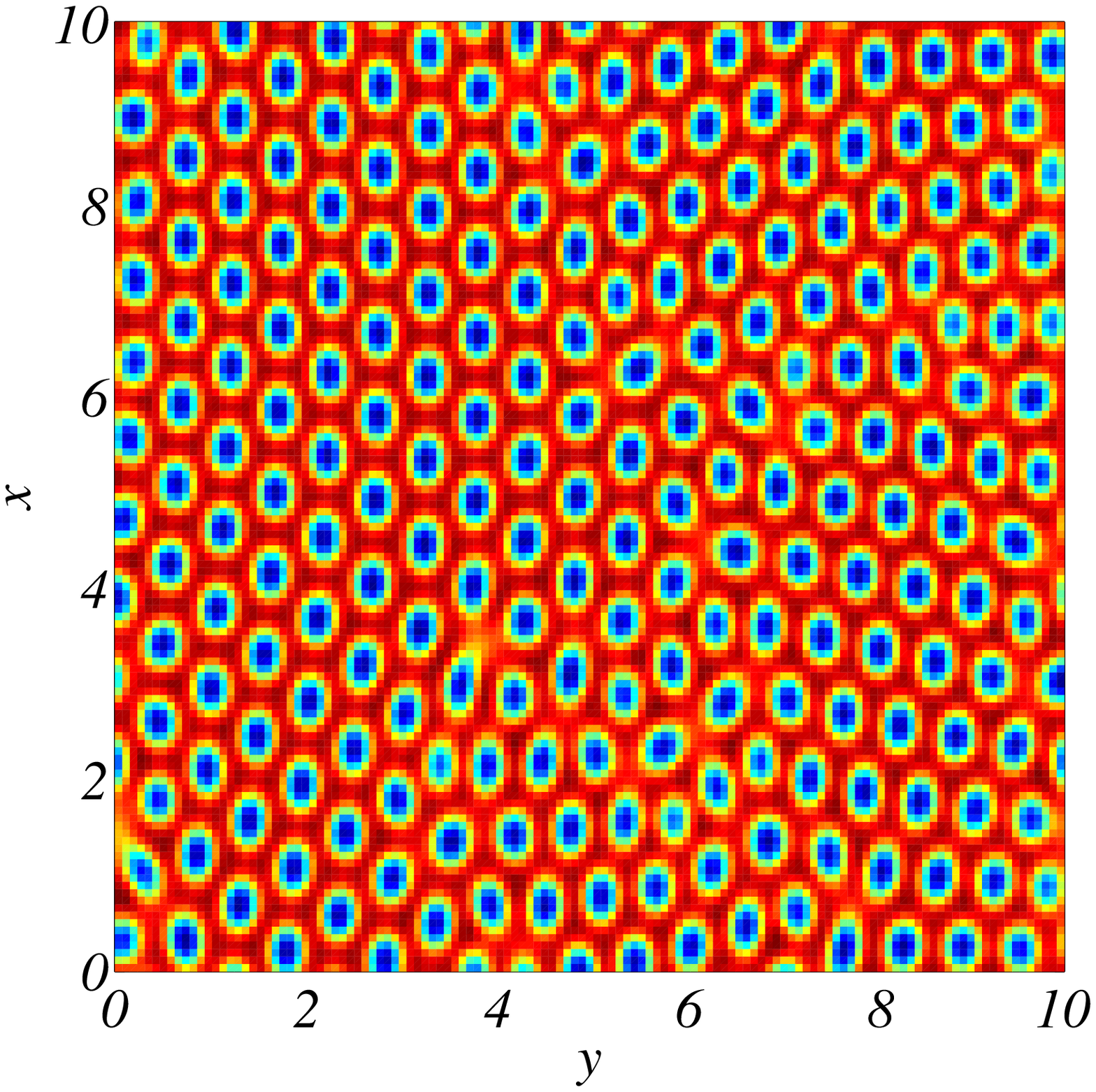}&
\includegraphics[width=6cm]{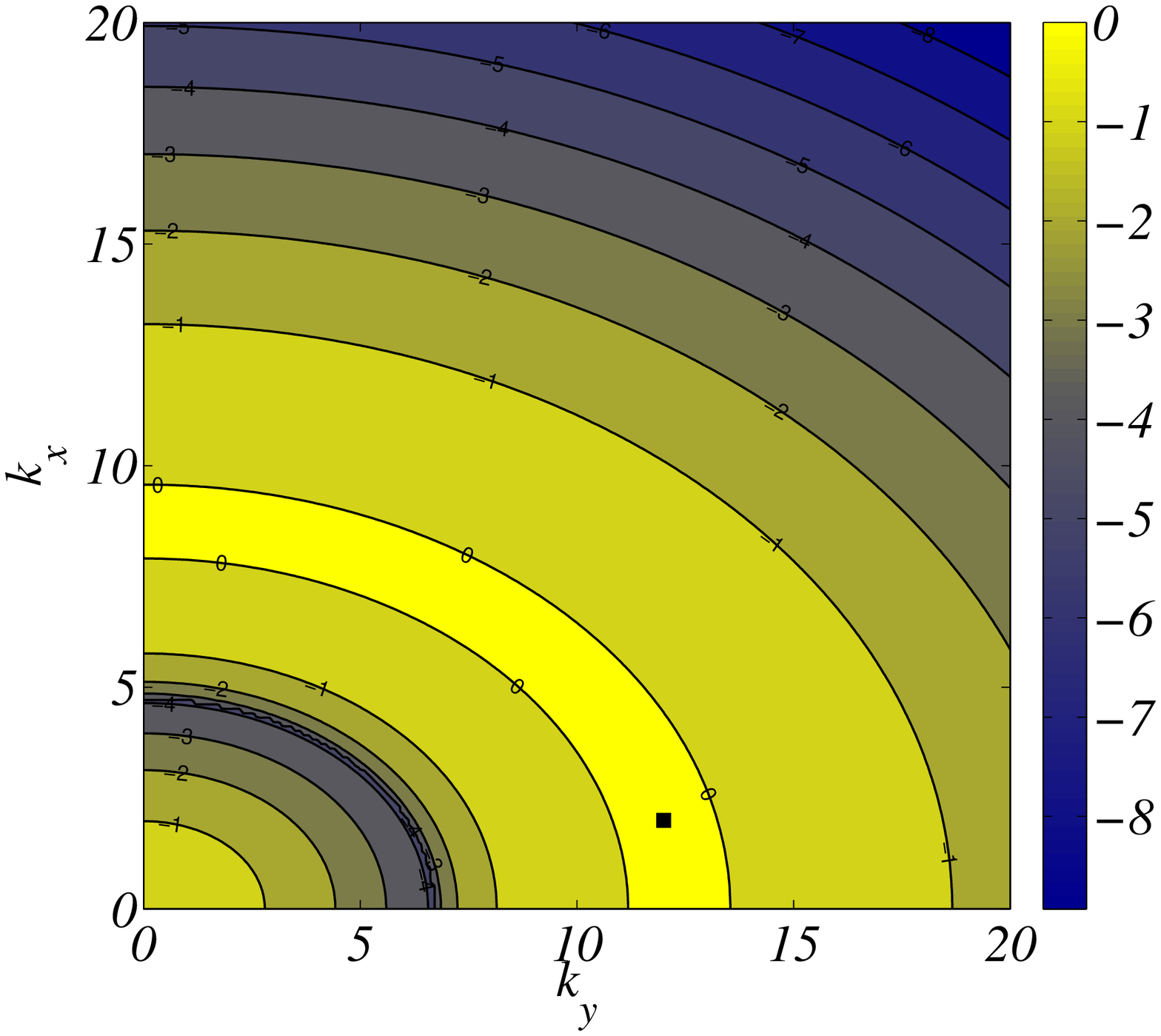}
\\
(a)& (b) \\
\includegraphics[width=6cm]{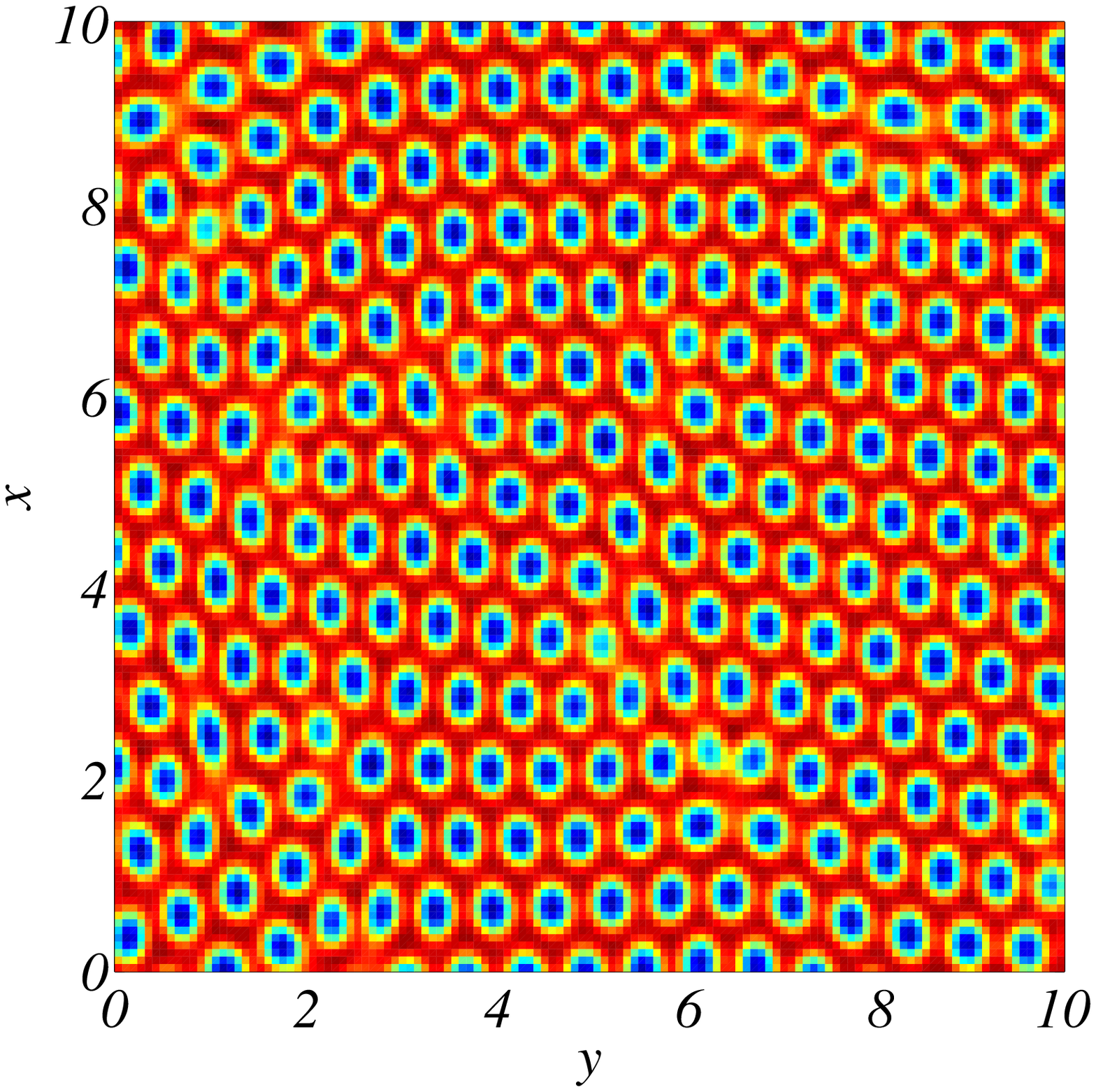}&
\includegraphics[width=6cm]{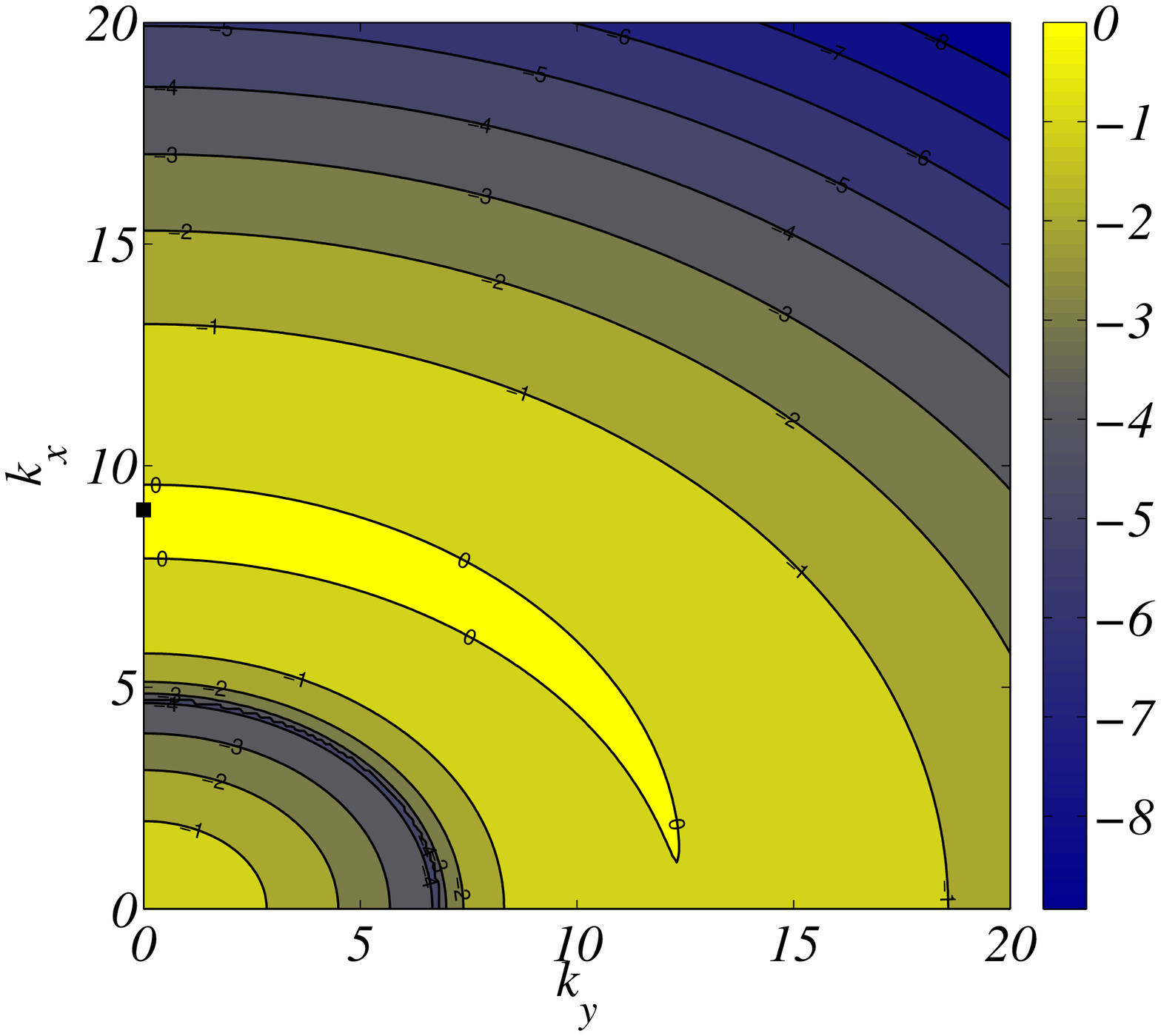}
\\
(c)& (d) \\
\includegraphics[width=6cm]{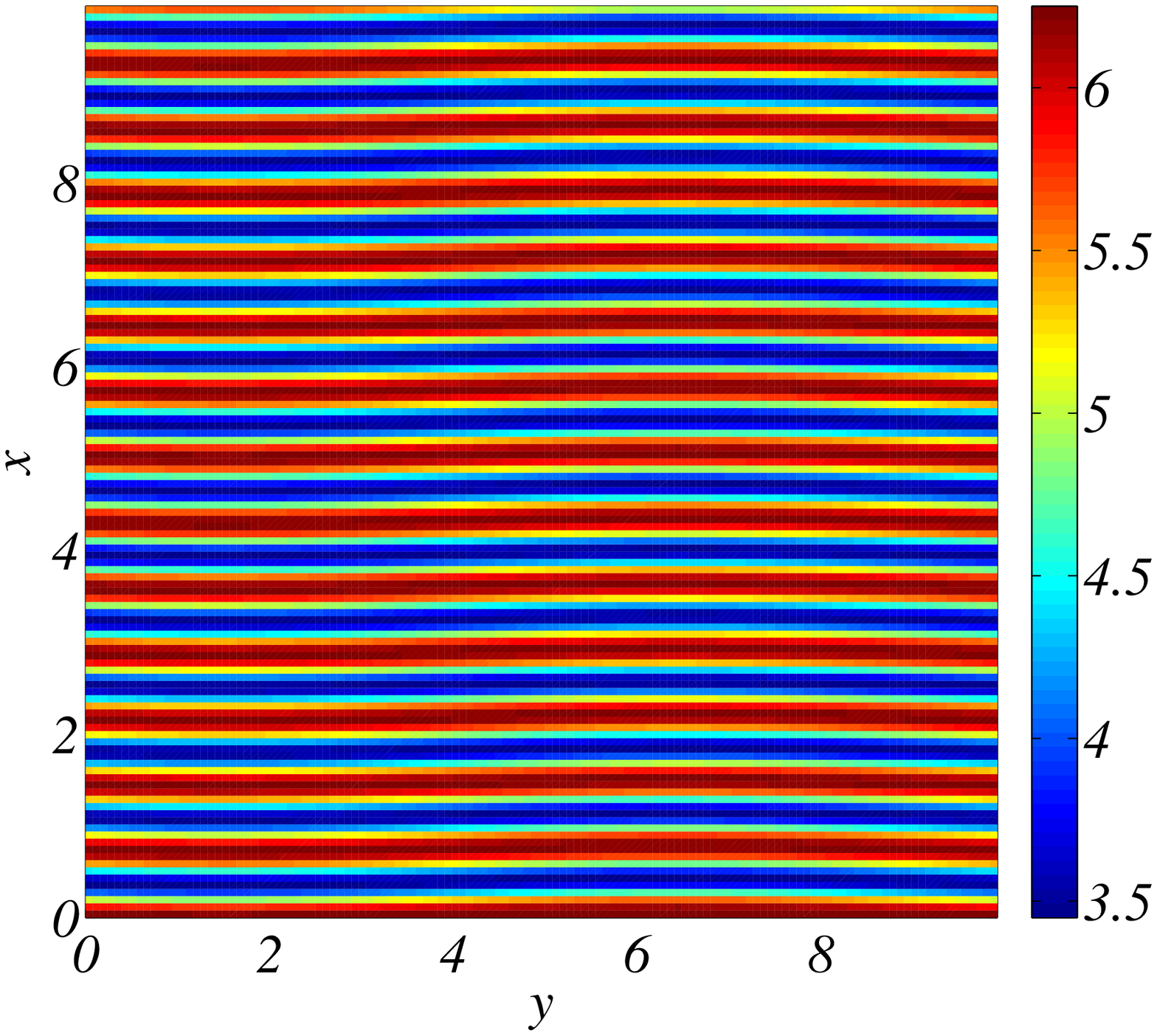}&
\includegraphics[width=6cm]{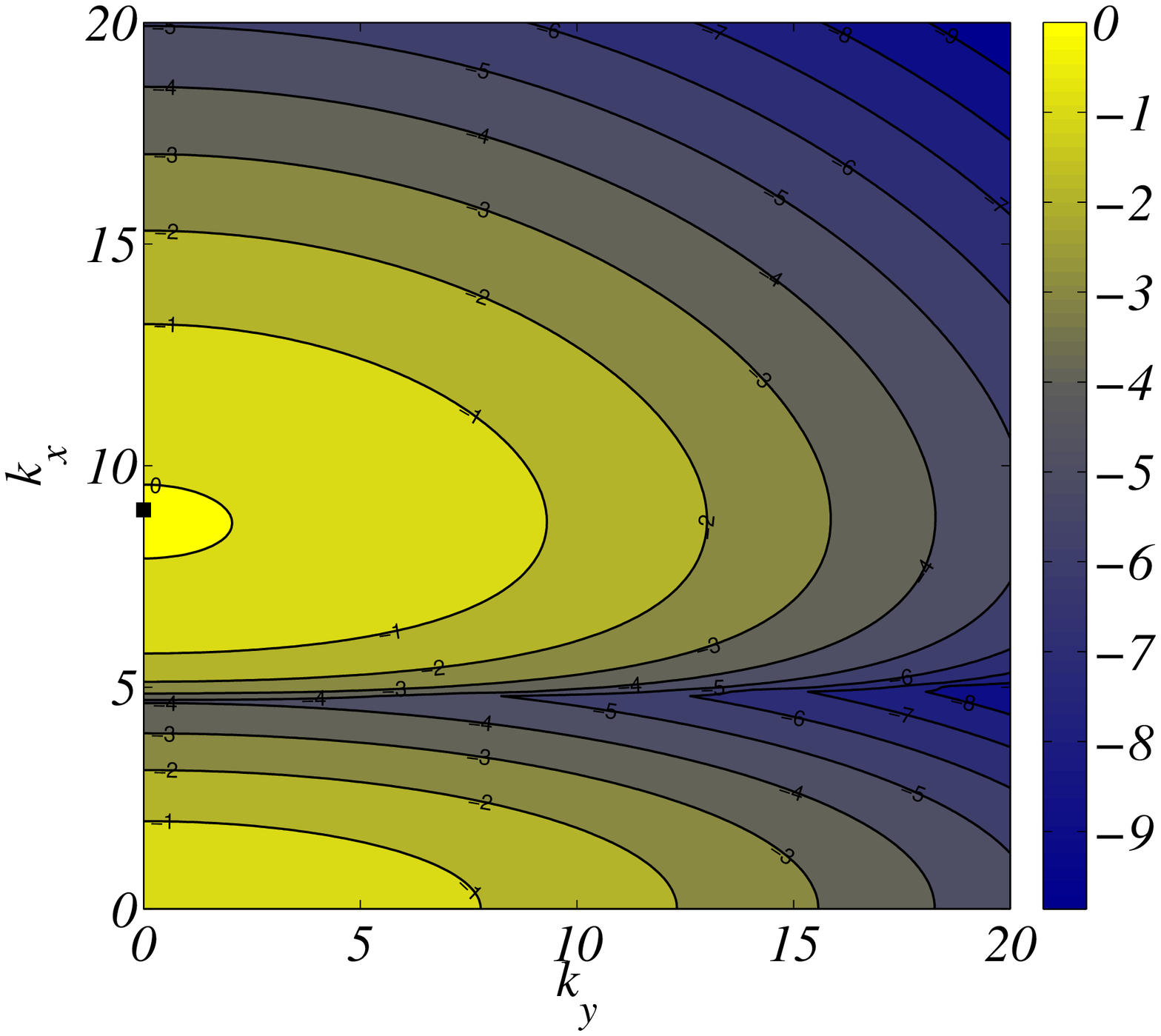}
\\
(e)& (f) \\
\end{tabular}
\caption{Asymptotic activator distribution: the concentration $u(x,y,t)$ is displayed for sufficiently large $t$. Panel (a): $D_u^{(y)} = 0.01$, $D_v^{(y)} = 0.16$, $D_u^{(x)} = 0.02$, $D_v^{(x)} = 0.32$. Panel (b): $D_u^{(y)} = 0.01$, $D_v^{(y)} = 0.155$, $D_u^{(x)} = 0.02$, $D_v^{(x)} = 0.32$. 
Panel (c): $D_u^{(y)} = 0.012$, $D_v^{(y)} = 0.01$, $D_u^{(x)} = 0.02$, $D_v^{(x)} = 0.32$. The other parameters are set to $a=35$; $b=16$;
$c=9$; $d=0.4$. The filled black squares identify the position of the maximum of the dispersion relation.
}
\label{fig:patterns}
\end{figure}

The aim of this section is to discuss a numerical implementation of the theory presented above. In particular, we will show that complex patterns can 
emerge for a system of two species in mutual interaction and undergoing anisotropic diffusion, also if the conventional Turing request of having inhibitors faster than activators is relaxed, along one of the two orthogonal directions of movements. To perform the analysis we operate in the framework of the so called Mimura-Murray model~\cite{MimuraMurray1978}. The quantities $u$ and $v$ can be associated to prey and predator densities, which interact via the non-linear functions:
\begin{equation}
\label{eq:fgMM}
f(u,v)= \left( (a+bu-u^2)/c - v\right) u \text{ and } g(u,v)= \left( u - (1+dv) \right) v\, ;
\end{equation}
the model possesses $6$ equilibria, whose stability and positivity depend on the value of the chosen parameters. We here focus on the fixed point ($\hat{u},\hat{v}$)
\begin{equation}
\label{eq:eqMM}
\hat{u}= 1+\frac{bd - 2d - c + \sqrt{\Delta}}{2d}\text{ and }\hat{v}= \frac{bd - 2d - c + \sqrt{\Delta}}{2d^2}\text{ where }\Delta = (bd-2d-c)^2 + 4d^2(a + b -1)\, ,
\end{equation}
and assume $a=35$, $b=16$, $c=9$ and $d=0.4$ which in turn implies $(\hat{u},\hat{v}) = (5,10)$. Moreover, the Jacobian entries evaluated at the fixed point reads
$f_u=3.33$, $f_v=-5$, $g_u=10$ and $g_v=-4$. Hence, $\det(J)>0$ and $\mathrm{tr}(J)<0$: the fixed point is a stable equilibrium. We also remark that $u$ acts as the activator and $v$ stands for the inhibitor species, as $f_u>0$ and $g_v<0$. Under specific conditions, the fixed point can be destabilized by an external, non homogeneous, perturbation, paving the way to the subsequent generation of Turing patterns, in the non linear regime of the evolution. In
Fig.~\ref{fig:patterns} we report a gallery of representative patterns that can be obtained under distinct conditions.

To generate the asymptotic patterns displayed in panel (a) of Fig~\ref{fig:patterns}, parameters are set  
so that both relations~\eqref{eq:case1} and~\eqref{eq:case2} are satisfied, {$D^{(x)}_v > D^{(x)}_ur_c$ and $D^{(y)}_v > D^{(y)}_ur_c$, where $r_c\sim 16$}. Inhibitor diffuses faster than activators in both $x$ and $y$ directions, although with different diffusion constants. The dispersion relation (see Fig~\ref{fig:patterns}(b)) can be assimilated to that sketched in Fig.~\ref{fig:hsign}(d), and the corresponding patterns share marked similarities with those obtained in the conventional case of isotropic transport. 

In panel (c) of Fig.~\ref{fig:patterns}, conditions~\eqref{eq:case1} hold, while~\eqref{eq:case2} do not, {$D^{(x)}_v > D^{(x)}_ur_c$ while $D^{(y)}_v < D^{(x)}_ur_c$, where $r_c\sim 16$}. The dispersion relation, Fig.~\ref{fig:patterns}(d), is also depicted and shown to resemble that displayed in Fig.~\ref{fig:hsign}(c). The patterns which follow this unusual choice of the diffusion constants, compare nicely with those emerging under the standard paradigm, {this is because $D^{(y)}_v / D^{(x)}_u$ is smaller but close to $r_c$}.

Finally, in panel (e,f) of Fig~\ref{fig:patterns},  the activator is assigned a diffusion coefficient  $D_u^{(y)}$ is larger than  $D_v^{(y)}$, the homologous constant associated to the inhibitor species, {and still $D^{(x)}_v > D^{(x)}_ur_c$}. The dispersion relation falls in the category exemplified in Fig.~\ref{fig:hsign}(c), and the corresponding patterns are found to organize in regular stripes, which run {almost} parallel to the direction where the instability is present.

\section{Conclusions}
\label{sec:concl}

In this paper we elaborated on the impact of anisotropic diffusion for the emergence of Turing patterns in reaction--diffusion systems. We have in particular focused on systems of two interacting species confined in a rectangular, continuum domain, endowed with periodic boundary conditions. With reference to this paradigmatic case study, we have shown that a symmetry breaking instability of the Turing type can occur only if patterns do exist   
when diffusion is impeded along one of the two accessible directions. In other words, patterns which resemble those obtained in the conventional setting of  isotropic diffusion emerge, also when the standard Turing condition is violated along one specific direction. Interestingly, the instability can also occur if the activator diffuses faster than the inhibitor, along the direction of spatial relocation for which the usual Turing conditions are not met.   

\section*{Acknowledgments}
The work of T.C. presents research results of the Belgian Network DYSCO (Dynamical Systems, Control, and Optimization), funded by the Interuniversity Attraction Poles Programme, initiated by the Belgian State, Science Policy Office.

\end{document}